# Effect of Vanadium Thickness and Deposition Temperature on VO$_2$ Synthesis using Atmospheric Pressure Thermal Oxidation

Ashok P, Yogesh Singh Chauhan, and Amit Verma

Department of Electrical Engineering, Indian Institute of Technology Kanpur, Kanpur 208016, India

Email: ashok@iitk.ac.in and amitkver@iitk.ac.in

ABSTRACT: Vanadium dioxide (VO$_2$) is a phase transition material that undergoes a reversible insulator-metal phase transition at ~ 68 ˚C. Atmospheric pressure thermal oxidation (APTO) of vanadium (V) is a simple VO$_2$ synthesis method in which V thin film is oxidized in open air. For an optimum oxidation duration, VO$_2$ films are obtained with good phase transition properties. We recently reported a modified APTO process using a step temperature profile for oxidation (Thin Solid Films 706, 138003 (2020)). We demonstrated an ultra-low thermal budget synthesis of VO$_2$ thin films with good electrical and optical phase transition properties. For a 130 nm room-temperature RF sputtered V thin film, an optimum oxidation duration of ~ 30 s was obtained. In this work, we study how the starting V film thickness and deposition temperature affects the optimum oxidation duration. V thin films of varying thickness (15-212 nm) and 120 nm thick V films with varying deposition temperature (~27-450 ˚C) are prepared using RF magnetron sputtering. These films are oxidized for different oxidation durations and characterized using Raman and four-probe measurements to find the optimum oxidation duration for each deposition condition. We find that the optimum oxidation duration increases with the increase in V film thickness and V deposition temperature. We model the effect of V film thickness and deposition temperature on the optimal oxidation time using a parabolic law which can be used to obtain the optimal oxidation times for intermediate V thicknesses/deposition temperatures.

1. Introduction

Vanadium dioxide (VO$_2$) is one of the most studied insulator to metal (IMT) phase transition material. It undergoes a first order structural phase transition from monoclinic insulator phase to tetragonal metallic phase above room temperature (~ 68 ˚C) [1]. The phase transition is accompanied with a significant change in resistivity [2] and infrared optical reflectance [3], because of which VO$_2$ finds use in many promising electronic and optical switching applications, such as radio frequency (RF) switches and other reconfigurable RF components [4, 5], coupled oscillators for neuromorphic computing [6], memresistors [7], selectors for resistive random-access memory [8], phase field-effect transistors [9, 10], microactuators [11], ultra-thin absorbers [12], tunable radiators [13], infrared camouflage [14], microbolometers [15], metamaterials [16], and thermochromic films [17].

In this work, atmospheric pressure thermal oxidation (APTO) of Vanadium (V) is used to synthesize VO$_2$ thin films. Unlike other VO$_2$ synthesis methods such as reactive sputtering [18, 19], chemical vapor deposition [20, 21], sol-gel synthesis [22], pulsed laser deposition [23, 24], reactive-evaporation [25], molecular beam epitaxy [26, 27], atomic layer deposition [28], and polymer assisted solution process [29], which need precise oxygen flow control, and/or high-temperature post-deposition annealing treatment (to stabilize $V^{+4}$ valency among multiple possible valence states of V [30]), APTO does not require controlled oxygen gas flow/post-synthesis annealing as it uses atmospheric oxygen for the oxidation of V [31-36]. APTO of V favors the formation of $VO_2$ during the initial oxidation duration. As oxidation proceeds, $VO_2$ content in the film maximizes at an optimal oxidation duration. Further oxidation leads to $V_2O_5$ formation and decrease in $VO_2$ content of the film [34, 36, 37]. Best phase transition properties (electrical/optical switching) are obtained at the optimum oxidation time ($t_{oxd}^{op}$) [32, 34, 36].

Recently, we modified the typical APTO process, by using a step temperature profile for oxidation, which allows us to precisely control the oxidation duration [36]. The thickness ($X_V$) and the deposition temperature ($T_{dep}$) of the V films can affect the time for their complete oxidation into $VO_2$. This dependence between the optimal oxidation duration and the V film thickness/deposition temperature has not been studied and is the subject of this work. We deposited V thin films with varying $X_V$ and at different $T_{dep}$ on c-plane Sapphire substrates using RF magnetron sputtering. We have performed APTO on the deposited samples and then characterized the oxidized samples to find the optimum oxidation time as a function of $X_V$ and $T_{dep}$. In section 2, we present the deposition conditions for the V thin films, the oxidation process, and the sample characterization methods. In section 3, Raman and electrical characterization results of the oxidized samples are presented along with the summarized results for the optimum oxidation durations as a function of $X_V$ and $T_{dep}$. Modeling and discussion of the optimum oxidation time results are presented in section 4. Conclusions of the study are presented in section 5.

## 2. Experimental Procedure

For preparing the thin film V samples on c-plane Sapphire substrates with varying $X_V$ and $T_{dep}$, we used RF magnetron sputtering from a 2-inch diameter V target of 99.9% purity. Base vacuum pressure was maintained in the range of 1.5 x $10^{-4}$- 8.7 x $10^{-4}$ Pa prior to the introduction of argon gas (purity of 99.999%) in the chamber. We use sputter down geometry and performed the V deposition at ~1 Pa, using a RF magnetron power of 90W which resulted in V deposition rate of ~0.5 Å/s. To study the film thickness effects during APTO, we prepared 4 samples with $X_V$ = 15 nm, 80 nm, 120 nm, and 212 nm, deposited at room temperature (R.T.) by varying the deposition duration. In addition to the $X_V$ = 120 nm V film deposited at R.T. for thickness series, we deposited $X_V$ = 120 nm films with varying $T_{dep}$ = 100 ˚C, 200 ˚C, 300 ˚C, 400 ˚C and, 450 ˚C to study the effects of V deposition temperature on APTO. Thickness, $X_V$, for all the samples was confirmed using a KLA-Tencor stylus profilometer.

After the depositions, all the samples were diced into multiple pieces to carry out the APTO experiments with varying oxidation time ($t_{oxd}$). After the dicing, the samples were cleaned using iso-propyl alcohol in a sonicator and blow-dried using dry nitrogen. For the oxidation experiments, we used a hot plate at 450 ˚C and a cold plate at room temperature for subjecting the sample to a step temperature profile, similar to our previous APTO work [36]. Diced V films of the $X_V$ and $T_{dep}$ series were oxidized for different $t_{oxd}$ in the 0-1800 s range to synthesize the vanadium oxide thin films.

The oxidized samples were then characterized by Raman spectroscopy using an excitation laser wavelength of 532 nm (Acton research corporation spectra pro 2500i) to determine the vanadium oxide phase of the films. Four-point probe resistance measurements with a temperature-controlled oven were performed to measure resistance switching properties of each sample.

## 3. Results

3.1 Raman and Electrical Characterization of V Thickness series

Fig. 1(a) shows Raman spectra of all the oxidized samples of $X_V$ = 212 nm as a function of the oxidation duration. Oxidized V films with $t_{oxd}$ = 95 s, 110 s, 120 s, 130 s, 140 s, and 150 s show strong $VO_2$ peaks [38]. Samples with $t_{oxd}$ =140 s and 150 s also show extremely weak $V_2O_5$ peaks at 148 $cm^{-1}$ and 995 $cm^{-1}$ [38]. Films with $t_{oxd}$ = 180 s, 240 s, and 300 s show mixed phases of $VO_2$ and $V_2O_5$ as three Raman peaks around 194 $cm^{-1}$, 226 $cm^{-1}$, and 613 $cm^{-1}$ relate to $VO_2$ and all other peaks are associated with $V_2O_5$. Films with $t_{oxd}$ = 600 s and 1800 s show only $V_2O_5$ Raman peaks. Raman spectrum of oxidized $X_V$ =120 nm thick V samples are shown in Fig. 1(b). Sample oxidized with $t_{oxd}$ = 70 s shows only $VO_2$ Raman peaks, while oxidized samples with $t_{oxd}$ = 77 s, 80 s, and 90 s show dominant $VO_2$ peaks and weak $V_2O_5$ peaks. The film with $t_{oxd}$ = 120 s shows mixed peaks of $VO_2$ and $V_2O_5$, while $t_{oxd}$ = 300 s, 600 s, and 1800 s show only $V_2O_5$ Raman peaks.

Fig. 1(c) shows the Raman spectrum as a function of the oxidation duration for $X_V$ = 80 nm thick V samples. Films oxidized with $t_{oxd}$ = 15 s, 20 s, and 21 s exhibit only $VO_2$ peaks. Film with $t_{oxd}$ = 22 s shows strong $VO_2$ peaks and weak $V_2O_5$ peaks. Samples with $t_{oxd}$ = 30 s and 45 s show mixed phases of $VO_2$ and $V_2O_5$, while $t_{oxd}$ = 600 s and 1800 s samples show only $V_2O_5$ Raman peaks. Fig. 1(d) shows Raman spectra of the thinnest oxidized samples with $X_V$ = 15 nm. Sample oxidized with $t_{oxd}$ = 3 s does not show any significant Raman peak while $t_{oxd}$ = 4 s sample shows only $VO_2$ peaks. $t_{oxd}$ = 5 s sample shows mixed Raman peaks of $VO_2$ and $V_2O_5$ while $t_{oxd}$ = 600 s and 1800 s show only $V_2O_5$ peaks. From the Raman characterization, we can conclude that for APTO of all the V thicknesses, $VO_2$ forms in the initial stages while $V_2O_5$ forms in the latter stages, in agreement with the previous studies [34, 36, 37].

To study the phase transition properties of the oxidized samples as a function of $t_{oxd}$, we measured the four-point probe resistance of all the samples at 30 ˚C and 110 ˚C. At 30 ˚C, $VO_2$ is expected to be in the insulating phase while at 110 ˚C, $VO_2$ should transition to the metallic phase changing the resistance drastically for samples with high $VO_2$ content [2, 32, 34, 36]. Fig. 2(a) shows the resistance switching ratio $R_{30 ˚C}/R_{110 ˚C}$ of oxidized $X_V$ = 212 nm samples as a function of $t_{oxd}$. Films oxidized with $t_{oxd}$ = 95 s and 110 s do not show any significant switching. From the Raman characterization, these samples show only $VO_2$ phase, but the absence of switching implies only partial oxidation of V. Samples with $t_{oxd}$ = 120 s, 130 s, and 140 s show strong resistance switching indicating minimal unoxidized V content in the films. Among these films, $t_{oxd}$ = 140 s sample shows the best switching ratio of more than three orders of magnitude. Fig. 2(b) shows the resistance as a function of temperature for sample with $t_{oxd}$ = 140 s. To extract the transition temperature during heating scan, hysteresis width, and transition width (during heating scan), we have fitted a Gaussian curve to the derivative of log10 R(T). The extracted transition temperature during heating scan, hysteresis width, and transition width (during heating scan) was found to be ~72.9 °C, ~9.8 °C, and ~7.1 °C, respectively. Films with $t_{oxd}$ =150 s, 180 s, 240 s, and 300 s show reduced IMT switching of more than two orders. The reduced switching is due to increasing $V_2O_5$ content in the films, as confirmed by Raman characterization. The switching order becomes less than one for $t_{oxd}$ = 600 s and 1800 s samples suggesting dominant $V_2O_5$ content in the samples. The results of Raman characterization and resistance switching suggest that $t_{oxd}$ = 140 s is the optimal oxidation time ($t_{oxd}^{op}$) for synthesizing $VO_2$ film with good IMT switching properties using APTO of $X_V$ = 212 nm V thin films.

The resistance switching ratio of $X_V$ = 120 nm oxidized samples are shown in Fig. 2(c). $t_{oxd}$ = 70 s film shows resistance switching of more than one order due to partial oxidation of V. Films with $t_{oxd}$ = 77 s and 80 s exhibit good resistance switching. The best resistance switching of more than three orders is achieved for $t_{oxd}$ = 80 s. Fig. 2(d) shows the temperature scan of resistance for sample with $t_{oxd}$ = 80 s. Flim with $t_{oxd}$ = 80 s shows reversible resistance switching with extracted transition temperature, hysteresis width, and transition width as be ~76.8 °C, ~12.6 °C, and ~9.3 °C, respectively. After $t_{oxd}$ =80 s, IMT switching reduces sharply. No significant switching is observed for $t_{oxd} \geq$ 300 s. Fig. 2(e) summarizes the resistance switching ratio of all the oxidized samples with $X_V$ = 80 nm as a function of $t_{oxd}$. The resistance switching increases monotonically for samples with $t_{oxd}$ = 15 s, 20 s, and 21 s. Among these samples, $t_{oxd}$ = 21 s sample exhibits the best switching of more than two orders. Fig. 2(f) shows reversible resistance switching of $t_{oxd}$ = 21 s sample. The extracted transition temperature, hysteresis width, and transition width was found to be ~71.7 °C, ~11.6 °C, and ~5.9 °C, respectively. After $t_{oxd}$ > 21 s, IMT switching reduces steeply. Fig. 2(g) shows the resistance switching ratio of oxidized $X_V$ = 15 nm samples. Sample oxidized with $t_{oxd}$ = 3 s shows more than one order of switching, while $t_{oxd}$ = 4 s shows the best switching of nearly three orders of magnitude shown in Fig. 2(h). The extracted transition temperature, hysteresis width, and transition width was found to be ~77.7 °C, ~13 °C, and ~7.9 °C, respectively. The resistance switching ratio reduces for further increase in $t_{oxd}$.

From both Raman and resistance switching characterization, we find that the optimum oxidation time, $t_{oxd}^{op}$, for achieving good switching $VO_2$ films is found to increase monotonically with increasing V thickness. Compared to $t_{oxd}^{op}$ = 4 s for $X_V$ = 15 nm V film, $X_V$ = 212 nm V film shows more than one order high $t_{oxd}^{op}$ = 140 s value.

3.2 Raman and Electrical Characterization of V Deposition Temperature Series

Raman data for APTO oxidation of V sample with $X_V$ = 120 nm film deposited at $T_{dep}$ = 27 °C (Fig. 1(b)) is repeated again in Fig. 3(a) for comparison with other samples of the $T_{dep}$ series. Fig. 3(b) shows the Raman spectrum as a function of $t_{oxd}$ of V films ($X_V$ = 120 nm) with $T_{dep}$=100 °C. Films oxidized with $t_{oxd}$ = 50 s and 60 s show only $VO_2$ Raman peaks while $t_{oxd}$ = 65 s shows dominant $VO_2$ peaks and some weak $V_2O_5$ peaks. Samples with $t_{oxd}$ = 70 s and 75 s show mixed phases of $VO_2$ and $V_2O_5$. $t_{oxd}$ = 80 s and 90 s show strong $V_2O_5$ Raman peaks and weak $VO_2$ peaks. $t_{oxd}$=1800 s sample shows only $V_2O_5$ Raman peaks. Strong $VO_2$ Raman peaks were observed for samples oxidized with $t_{oxd}$ = 50 s-65 s of V film with $T_{dep}$=100 ˚C. Raman spectrum of all the oxidized samples of V films with $T_{dep}$ =200 ˚C and $X_V$ = 120 nm are shown in Fig. 3(c). V film oxidized with $t_{oxd}$ = 85 s shows only $VO_2$ Raman peaks. Films with $t_{oxd}$ = 90 s, 95 s, and 100 s show strong $VO_2$ peaks and weak $V_2O_5$ peaks. $t_{oxd}$ = 110 s shows mixed phase of $VO_2$ and $V_2O_5$ while $t_{oxd}$ = 1800 s show only $V_2O_5$ Raman peaks.

Fig. 3(d) shows the Raman spectrum of V films with $T_{dep}$ = 300 ˚C as a function of $t_{oxd}$. Samples oxidized with $t_{oxd}$ = 110 s, 120 s, and 125 s show dominant $VO_2$ peaks with weak $V_2O_5$ peaks. $t_{oxd}$ = 130 s and 140 s show mixed phases of $VO_2$ and $V_2O_5$ while $t_{oxd}$ = 1800 s show only $V_2O_5$ peaks. Fig. 3(e) shows the Raman spectrum of all oxidized samples of V film with $T_{dep}$ = 400 ˚C. Films oxidized with $t_{oxd}$ = 100 s, 115 s, 125 s, and 130 s show strong $VO_2$ peaks. However, $t_{oxd}$ = 130 s also shows a weak $V_2O_5$ Raman peak around 148 $cm^{-1}$. $t_{oxd}$ = 140 s and 160 s samples show mixed phases of $VO_2$ and $V_2O_5$. The longer $t_{oxd}$ of 1800 s shows only $V_2O_5$ Raman peaks. Fig. 3(f) summarizes the Raman spectrum as a function of $t_{oxd}$ of V films deposited at $T_{dep}$ = 450 ˚C. V films oxidized with $t_{oxd}$ = 165 s, 170 s, show only $VO_2$ Raman peaks. Films with $t_{oxd}$ = 175 s show dominant $VO_2$ peaks and a weak $V_2O_5$ peak around 148$cm^{-1}$. $t_{oxd}$ = 190 s and 210 s samples show mixed phases of $VO_2$ and $V_2O_5$ while $t_{oxd}$ = 1800 s show only $V_2O_5$ Raman peaks.

Similar to the thickness series, the phase transition properties of oxidized samples in $T_{dep}$ series were also studied by the four-point probe resistance measurements at 30 ˚C and 110 ˚C. For comparison with other samples of the $T_{dep}$ series, Fig. 4(a) repeats the data (Fig. 2(b)) for resistance switching ratio $R_{30 ˚C}/R_{110 ˚C}$ of oxidized V films ($X_V$ = 120 nm) with $T_{dep}$ = 27 ˚C as a function of $t_{oxd}$. This data has already been discussed in the previous sub-section. Fig. 4(b) summarizes the resistance switching ratio of all oxidized V films with $T_{dep}$ =100 ˚C. Sample oxidized with $t_{oxd}$ = 50 s shows one order switching only despite Raman characterization showing only $VO_2$ peaks. This reduced switching is due to partial oxidation of the V film. Films with $t_{oxd}$ =60 s and 65 s show good IMT switching. Among these films, $t_{oxd}$ = 65 s shows the best switching of more than three orders of magnitude. Films with $t_{oxd}$ = 70 s, 75 s, and 80 s show reduced switching of more than two orders. After $t_{oxd}$ =80 s, switching further reduces due to increasing $V_2O_5$ content in the films. The switching almost disappears for $t_{oxd}$ = 1800 s indicating only $V_2O_5$ content in the sample in agreement with the Raman characterization.

The resistance switching of all oxidized V films with $T_{dep}$ = 200 ˚C is shown in Fig. 4(c). Samples oxidized with $t_{oxd}$ = 85 s, 90 s, 95 s, and 100 s all show good switching greater than two orders. The best switching of three orders is however observed for $t_{oxd}$ = 95 s sample, the switching reduces after this optimum oxidation time. The longer $t_{oxd}$ of 1800 s shows almost no IMT switching. These results are in agreement with the Raman characterization results of these samples. Fig. 4(d) shows the resistance switching as a function of $t_{oxd}$ of V films with $T_{dep}$ =300 ˚C. The switching ratio increases from $t_{oxd}$ = 110 s and reaches peak value at $t_{oxd}$ = 125 s. For $t_{oxd} \geq$ 130 s switching reduces and almost disappears at $t_{oxd}$ = 1800 s. Fig. 4(e) shows the resistance switching ratio of all oxidized V films with $T_{dep}$ = 400 ˚C. No significant switching is observed for $t_{oxd}$ = 100 s due to only partial oxidation of the V film. More than two orders of switching is observed for $t_{oxd}$ = 115-160 s. $t_{oxd}$ = 130 s sample shows the best switching of three orders of magnitude. Beyond this optimum oxidation time, the resistance switching ratio decreases and becomes small at $t_{oxd}$ = 1800 s. Fig.

4(f) summarizes the resistance switching as a function of $t_{oxd}$ for V film with $T_{dep}$ = 450 °C. Sample oxidized with $t_{oxd}$ = 165 s shows more than one order of switching. Films with $t_{oxd}$ = 170 s-210 s show good switching of more than two orders of magnitude. The peaks switching is observed for $t_{oxd}$ = 175 s after which the switching performance decreases because of the increasing $V_2O_5$ content of the samples. The optimum oxidation time $t_{oxd}^{op}$ of $T_{dep}$ = 450 °C sample is nearly three times more than $t_{oxd}^{op}$ of V with $T_{dep}$ = 100 °C even though both samples have the same V thickness.

From Raman and electrical characterization discussed above, we can conclude that the optimum oxidation duration, $t_{oxd}^{op}$, increases with V film thickness and with increase in V deposition temperature. Table 1 summarizes the optimal oxidation time ($t_{oxd}^{op}$) and the corresponding resistance switching ratio ($R_{30\ °C}/R_{110\ °C}$) as a function of V film thickness and deposition conditions. To confirm that $VO_2$ is indeed the dominant phase in the samples oxidized for $t_{oxd}^{op}$, Raman peak intensity ratio of $V_2O_5$ (~995 cm$^{-1}$) and $VO_2$ (~613 cm$^{-1}$) peaks is also listed in Table 1. This Raman intensity ratio is small for all the $t_{oxd}^{op}$ samples confirming dominant $VO_2$ and extremely weak $V_2O_5$ presence.

4. **Discussion**

Room temperature and high temperature resistivity of $VO_2$ films synthesized from V films deposited at different deposition temperatures ($T_{dep}$) is shown in Fig.5. For comparison, room temperature and high temperature resistivity of $VO_2$ films deposited using reactive sputtering (V- target & Ar+air) at 450 °C and RF sputtering (from $VO_2$- target) at 550 °C [39] is also shown. The resistivity $\rho$ is extracted from the four probe measurements. The probe separation in the four-probe setup is 1.5 mm which is much larger than the $VO_2$ film thickness allowing the use of following equation for the calculation of resistivity [40],

$$\rho = \frac{\pi t}{\ln 2} \frac{V}{I} \quad (1)$$

where $I$ is the current supplied through the outer terminals of four-probe, $V$ is the voltage measured between the inner two probes and $t$ is the film thickness. Vanadium dioxide thickness $t$ is taken to be twice of the thickness of Vanadium [31]. Resistivity values for $VO_2$ films synthesized using APTO of V lies between the resistivity values of the $VO_2$ synthesized from reactive sputtering of V and direct sputtering of $VO_2$. As $T_{dep}$ increases APTO $VO_2$ film resistivity approaches the resistivity of $VO_2$ films synthesized using RF sputtering (from $VO_2$- target) at 550 °C.

Optimum oxidation duration ($t_{oxd}^{op}$) as a function of the V film thickness ($X_V$) deposited at R.T. is shown in Fig. 6 (a). Clearly, $t_{oxd}^{op}$ increases monotonically with increase in $X_V$. As more time is needed for oxygen to diffuse into a thicker V film to convert it into $VO_2$, the trend in $t_{oxd}^{op}$ is understandable. To quantitatively understand the variation, we model the data using a parabolic fitting expression, $X_V^2 = Bt_{oxd}^{op}$, where B is the parabolic constant. Parabolic fit matches with the experimentally obtained $t_{oxd}^{op}$ with different vanadium thicknesses (B = 192.9 nm$^2$/s) except the thicker film with $X_V$=212 nm, suggesting different growth law for thicker films. As V completely converts into $VO_2$ at $t_{oxd} = t_{oxd}^{op}$ and in this oxidation process the film thickness increases by ~2-2.1 times [31], we can conclude that $VO_2$ thickness vs oxidation time will also follow a parabolic growth law with a parabolic constant of 771.6-850.6 $nm^2/s$.

Fig. 6(b) shows $t_{oxd}^{op}$ with increasing V deposition temperature ($T_{dep}$) for $X_V$ = 120 nm thick V films. The optimum oxidation time in general shows an increasing trend with the increasing $T_{dep}$. The deposition temperature is expected to affect the microstructural properties of V films and hence the oxygen diffusion rates. This hypothesis is supported by Fig. 6(c), which shows the conductivity of the deposited V samples as a function of $T_{dep}$. The conductivity is found to be nearly same in the low deposition temperature regime (~27-200 °C), beyond which it increases exponentially suggesting that microstructural changes are significant beyond ~ 200 °C. We model the V conductivity as a function of $T_{dep}$ using the following mathematical expression,

$$\sigma = \sigma_0 e^{\frac{-E_{a1}}{kT}} + \sigma_c \qquad (2)$$

where, the $\sigma_c$ term models the saturation behaviour in the low temperature regime (~27-200 °C), while the exponential term models the exponential rise of V conductivity in the high temperature. For the values shown in the inset of Fig. 6(c), this expression fits the measured V conductivity vs $T_{dep}$ data reasonably well. This activation energy is possibly from the contribution of grain growth in V films (with $T_{dep}$) to the improvement in electron transport/conductivity in the V films. All samples in the $T_{dep}$ series have the same thickness $X_V$ = 120 nm so from the parabolic relation used above, $X_V^2 = B t_{oxd}^{op}$, we can conclude that any increase in $t_{oxd}^{op}$ should be compensated with a corresponding decrease in the parabolic constant B as a function of deposition temperature. We use this inverse relation to find deposition temperature dependence of 1/B as shown in Fig. 6(d). As the behavior of 1/B vs $T_{dep}$ (Fig. 6(d)) is similar to the behavior of V conductivity vs $T_{dep}$ (Fig. 6(c)), we use similar mathematical expression to fit the 1/B vs $T_{dep}$ dependence as shown in Fig. 6(d). The extracted activation energy in this case is possibly from the contribution of V grain growth/decrease in grain boundaries (with $T_{dep}$) towards decrease in oxygen diffusion rates during the APTO process.

## 5. Conclusion

In conclusion, we have experimentally studied the effect of V film thickness and V deposition temperature, on the optimum oxidation time for obtaining $VO_2$ films with best resistance switching properties, during the APTO process. From the Raman and electrical characterization performed on the synthesized samples, it was found that optimal oxidation time for synthesizing $VO_2$ films increases monotonically with increase in V film thickness, following a parabolic oxidation law. The optimal oxidation time was found to increase with the V deposition temperature. We explain and model this behavior by establishing a correlation with V conductivity variation with the deposition temperature. This study provides an insight into the effect of two major factors involved in $VO_2$ synthesis using APTO and should lead to wider use of this technique for realizing $VO_2$ based devices.

Acknowledgements:

This project was supported by IIT Kanpur initiation grant and used Materials Science and Engineering (MSE) Raman characterization facility.

**REFERENCES**

[1] J. B. Goodenough, The Two Components of the Crystallographic Transition in $VO_2$, J. Solid State Chem. 3 (1971) 490–500.

[2] Larry A. Ladd, William Paul, Optical and transport properties of high quality crystals of $V_2O_4$ near the metallic transition temperature, Solid State Commun. 7 (1969) 425-428.

[3] A.S. Barker Jr., H.W. Verleur, H.J. Guggenheim, Infrared optical properties of vanadium dioxide above and below the transition temperature, Phys. Rev. Lett., 17 (1966) 1286-1289.

[4] F. Dumas-Bouchiat, C. Champeaux, A. Catherinot, A. Crunteanu, P. Blondy, RF-microwave switches based on reversible semiconductor-metal transition of $VO_2$ thin-films synthesized by pulsed-laser deposition, Appl. Phys. Lett. 91 (2007) 223505.

[5] S. D. Ha, Y. Zhou, A. E. Duwel, D. W. White, S. Ramanathan, Quick Switch: Strongly Correlated Electronic Phase Transition Systems for Cutting-Edge Microwave Devices, in IEEE Microwave Magazine 15 (2014) 32-44.

[6] Nikhil Shukla, Abhinav Parihar, Eugene Freeman, Hanjong Paik, Greg Stone, Vijaykrishnan Narayanan, Haidan Wen, Zhonghou Cai, Venkatraman Gopalan, Roman Engel-Herbert, Darrell G. Schlom, Arijit Raychowdhury, Suman Datta, Synchronized charge oscillations in correlated electron systems, Sci Rep 4, (2015) 4964.

[7] T. Driscoll, H.T. Kim, B.G. Chae, M. Di Ventra, D.N. Basov, Phase-transition driven memristive system, Appl. Phys. Lett. 95 (2009) 043503.

[8] I. Radu, B. Govoreanu, K. Martens, Vanadium Dioxide for Selector Applications, ECS Trans. 58 (2013) 249–258.

[9] N. Shukla, A. V Thathachary, A. Agrawal, H. Paik, A. Aziz, D. G. Schlom, S. K. Gupta, R. Engel-Herbert, S. Datta, A steep-slope transistor based on abrupt electronic phase transition, Nat. Commun. 6 (2015) 8475.

[10] A. Verma, B. Song, B. Downey, V.D. Wheeler, D. J. Meyer, H. G. Xing, D. Jena, Steep Sub-Boltzmann Switching in AlGaN/GaN Phase-FETs with ALD $VO_2$, IEEE Trans. Electron Devices. 65 (2018) 945–949.

[11] K. Liu, S. Lee, S. Yang, O. Delaire, J. Wu, Recent progresses on physics and applications of vanadium dioxide, Mater. Today. 21 (2018) 875-896.

[12] M.A. Kats, D. Sharma, J. Lin, P. Genevet, R. Blanchard, Z. Yang, M.M. Qazilbash, D.N. Basov, S. Ramanathan, F. Capasso, Ultra-thin perfect absorber employing a tunable phase change material, Appl. Phys. Lett. 101 (2012) 221101.

[13] M. Benkahoul, M. Chaker, J. Margot, E. Haddad, R. Kruzelecky, B. Wong, W. Jamroz, P. Poinas, Thermochromic $VO_2$ film deposited on Al with tunable thermal emissivity for space applications, Sol. Energy Mater. Sol. Cells. 95 (2011) 3504-3508.

[14] L. Xiao, H. Ma, J. Liu, W. Zhao, Y. Jia, Q. Zhao, K. Liu, Y. Wu, Y. Wei, S. Fan, Fast adaptive thermal camouflage based on flexible $VO_2$/Graphene/CNT thin films, Nano Lett. 15 (2015) 8365-8370.

[15] C. Chen, X. Yi, X. Zhao, B. Xiong, Characterizations of VO$_2$ based uncooled microbolometer linear array Sens. Actuators, A, 90 (2001) 212-214.

[16] Jitendra K. Pradhan, S. Anantha Ramakrishna, Bharathi Rajeswaran, Arun M. Umarji, Venu Gopal Achanta, Amit K. Agarwal, Amitava Ghosh, High contrast switchability of VO$_2$ based metamaterial absorbers with ITO ground plane, Optics Express, 25 (2017) 9116.

[17] Y. Cui, Y. Ke, C. Liu, Z. Chen, N. Wang, L. Zhang, Y. Zhou, S. Wang, Y. Gao, Y. Long, Thermochromic VO$_2$ for energy-efficient smart windows, Joule 2 (2018) 1-40.

[18] E. Kusano, J. A. Theil, J.A. Thornton, Deposition of vanadium oxide films by direct-current magnetron reactive sputtering, J. Vac. Sci. Technol. A. 6 (1988) 1663–1667.

[19] D. Ruzmetov, S. D. Senanayake, V. Narayanamurti, S. Ramanathan, Correlation between metal-insulator transition characteristics and electronic structure changes in vanadium oxide thin films, Phys. Rev. B. 77 (2008) 195442.

[20] H. Zhane, H.L.M. Chans, J. Guo, T.J. Zhane, Microstructure of epitaxial VO$_2$ thin films deposited on (1120) sapphire by MOCVD, J. Mater. Res. 9 (1994) 2264–2271.

[21] M.B. Sahana, G.N. Subbanna, S.A. Shivashankar, Phase transformation and semiconductor-metal transition in thin films of VO$_2$ deposited by low-pressure metalorganic chemical vapor deposition, J. Appl. Phys. 92 (2002) 6495–6504.

[22] B. Chae, H. Kim, S. Yun, B. Kim, Y. Lee, D. Youn, K. Kang, Highly oriented VO$_2$ thin films prepared by Sol-Gel deposition, Electrochem. Solid-State Lett. 9 (2006) C12–C14.

[23] D.H. Kim, H.S. Kwok, Pulsed laser deposition of VO$_2$ thin films, Appl. Phys. Lett. 65 (1994) 3188–3190.

[24] T. H. Yang, R. Aggarwal, A. Gupta, H. Zhou, R.J. Narayan, J. Narayan, Semiconductor-metal transition characteristics of VO$_2$ thin films grown on c-and r-sapphire substrates, J. Appl. Phys. 107 (2010) 053514.

[25] F. C. Case, Low temperature deposition of VO$_2$ thin films, J. Vac. Sci. Technol. A. 8 (1990) 1395.

[26] H. Paik, J. A. Moyer, T. Spila, J.W. Tashman, J. A. Mundy, E. Freeman, N. Shukla, J. M. Lapano, R. Engel-Herbert, W. Zander, J. Schubert, D. A. Muller, S. Datta, P. Schiffer, D.G. Schlom, Transport properties of ultra-thin VO$_2$ films on (001) TiO$_2$ grown by reactive molecular-beam epitaxy, Appl. Phys. Lett. 107 (2015) 163101.

[27] L. L. Fan, S. Chen, Y. F. Wu, F. H. Chen, W. S. Chu, X. Chen, C. W. Zou, Z. Y. Wu, Growth and phase transition characteristics of pure M-phase VO$_2$ epitaxial film prepared by oxide molecular beam epitaxy, Appl. Phys. Lett. 103 (2013) 131914.

[28] M. J. Tadjer, V. D. Wheeler, B. P. Downey, Z. R. Robinson, D. J. Meyer, C. R. Eddy, F. J. Kub, Temperature and electric field induced metal-insulator transition in atomic layer deposited VO$_2$ thin films, Solid. State. Electron. 136 (2017) 30–35.

[29] L. Kang, Y. Gao, H. Luo, A novel solution process for the synthesis of VO$_2$ thin films with excellent thermochromic properties, ACS Appl. Mater. Interfaces, 1 (2009) 2211-2218.

[30] Y. Kang, Critical evaluation and thermodynamic optimization of the VO–VO$_{2.5}$ system, J. Eur. Ceram. Soc. 32 (2012) 3187–3198.

[31] M. Gurvitch, S. Luryi, A. Polyakov, A. Shabalov, M. Dudley, G. Wang, S. Ge, V. Yakovlev, VO$_2$ films with strong semiconductor to metal phase transition prepared by the precursor oxidation process, J. Appl. Phys., 102 (2007) 033504.

[32] X. Xu, A. Yin, X. Du, J. Wang, J. Liu, X. He, X. Liu, Y. Huan, A novel sputtering oxidation coupling (SOC) method to fabricate VO$_2$ thin film, Appl. Surf. Sci. 256 (2010) 2750–2753.

[33] X. Xu, X. He, G. Wang, X. Yuan, X. Liu, H. Huang, S. Yao, H. Xing, X. Chen, and J. Chu, The study of optimal oxidation time and different temperatures for high quality VO$_2$ thin film based on the sputtering oxidation coupling method, Appl. Surf. Sci. 257 (2011) 8824–8827.

[34] G. Rampelberg, B. De Schutter, W. Devulder, K. Martens, I. Radu, C. Detavernier, In situ X-ray diffraction study of the controlled oxidation and reduction in the V–O system for the synthesis of VO$_2$ and V$_2$O$_3$ thin films, J. Mater. Chem. C. 3 (2015) 11357–11365.

[35] Sang, J.Wang, Y. Meng, X. Xu, J.L. Sun, Y.Wang, Z. Hua, T.Zheng, Z.Liu, C.Wang, C.Wu, Simple Method Preparation for Ultrathin VO$_2$ Thin Film and Control: Nanoparticle Morphology and Optical Transmittance, Jpn. J. Appl. Phys. 58 (2019) 050917.

[36] Ashok P, Y. S. Chauhan, Amit Verma, Vanadium Dioxide Thin Films Synthesized Using Low Thermal Budget Atmospheric Oxidation, Thin Solid Films 706 (2020) 138003.

[37] A. Mukherjee, S. P. Wach, Kinetics of the oxidation of vanadium in the temperature range 350-950 ºC, J. Less-Common Met. 92 (1983) 289–300.

[38] P. Shvets, O. Dikaya, K. Maksimova, A. Goikhman, A review of Raman spectroscopy of vanadium oxides,"J. Raman Spectrosc. 15 (2019) 1226-1244.

[39] D. Ruzmetov, K.T. Zawilski, V. Narayanamurti, S. Ramanathan, Structure-functional property relationships in rf-sputtered vanadium dioxide thin films, J. Appl. Phys., 102 (2007), 113715.

[40] I. Miccoli, F. Edler, H. Pfnür, and C. Tegenkamp, "The 100th anniversary of the four-point probe technique: the role of probe geometries in isotropic and anisotropic systems," J. Phys. Condens. Matter 27 (2015), 223201.

Table 1: Summary of the optimal oxidation time and the corresponding resistance switching ratio, and Raman peak intensity ratio obtained as a function of $X_v$ and deposition conditions. (R.T. = 27 ˚C)

| Vanadium Thickness (nm) $X_v$ | Vanadium Deposition Temperature $T_{dep}$ | Optimal Oxidation Time (s) $t_{oxd}^{op}$ | Resistance Switching $R_{30°C}/R_{110°C}$ Ratio | Raman Peak $V_2O_5$ (~995 cm$^{-1}$)/$VO_2$ (~613 cm$^{-1}$) Intensity Ratio |
|---|---|---|---|---|
| 212 | R.T. | 140 | 1520 | 0.05 |
| 120 | R.T. | 80 | 1005 | 0.06 |
| 80 | R.T. | 21 | 300 | ~0 |
| 15 | R.T. | 4 | 605 | ~0 |
| 120 | 100 ˚C | 65 | 1404 | 0.05 |
| 120 | 200 ˚C | 95 | 1225 | 0.06 |
| 120 | 300 ˚C | 125 | 357 | 0.19 |
| 120 | 400 ˚C | 130 | 1000 | ~0 |
| 120 | 450 ˚C | 175 | 446 | ~0 |

**FIGURES**

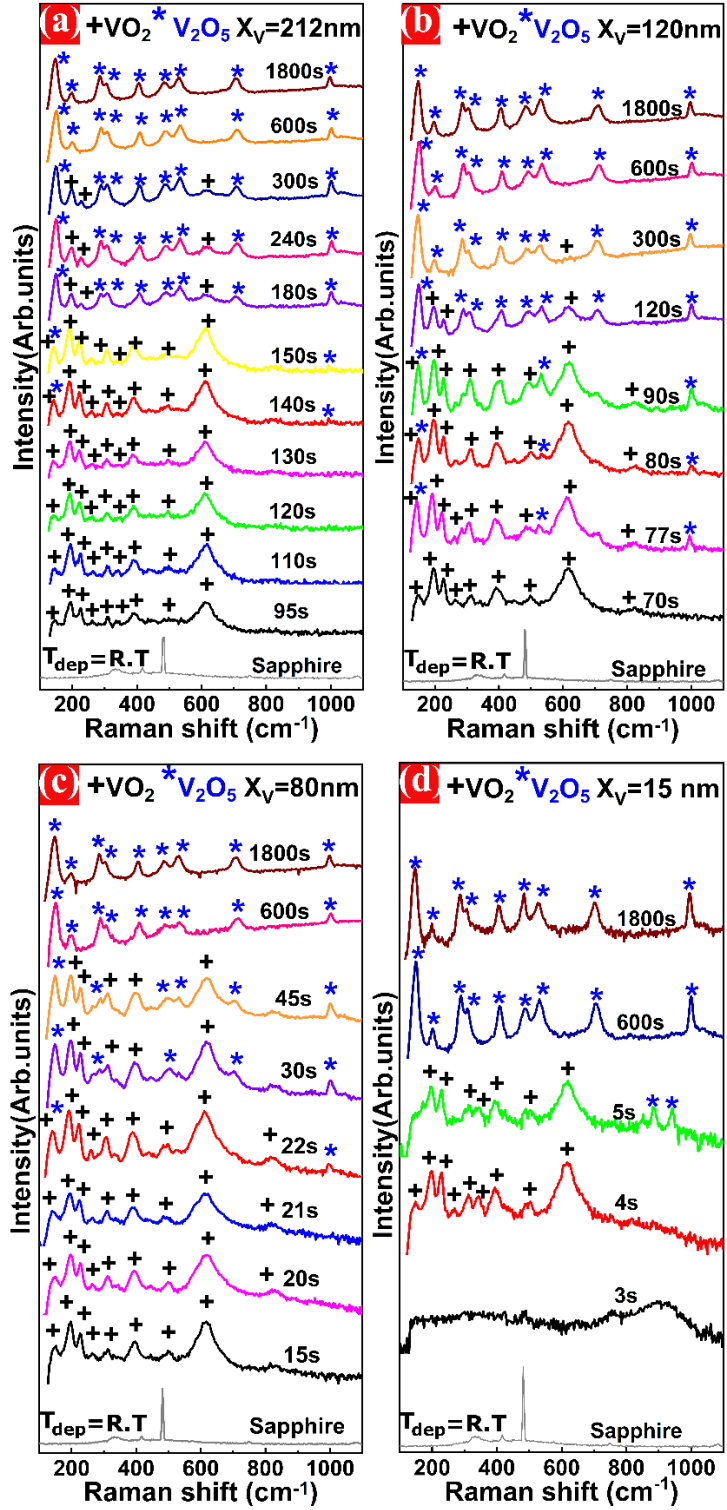

FIG. 1. Raman spectra of all oxidized Vanadium films with different thickness $X_v$, (a) 212 nm, (b) 120 nm, (c) 80 nm, and (d) 15 nm. Reference values of the Raman peak positions for $VO_2$ and $V_2O_5$ are taken from ref. [38].

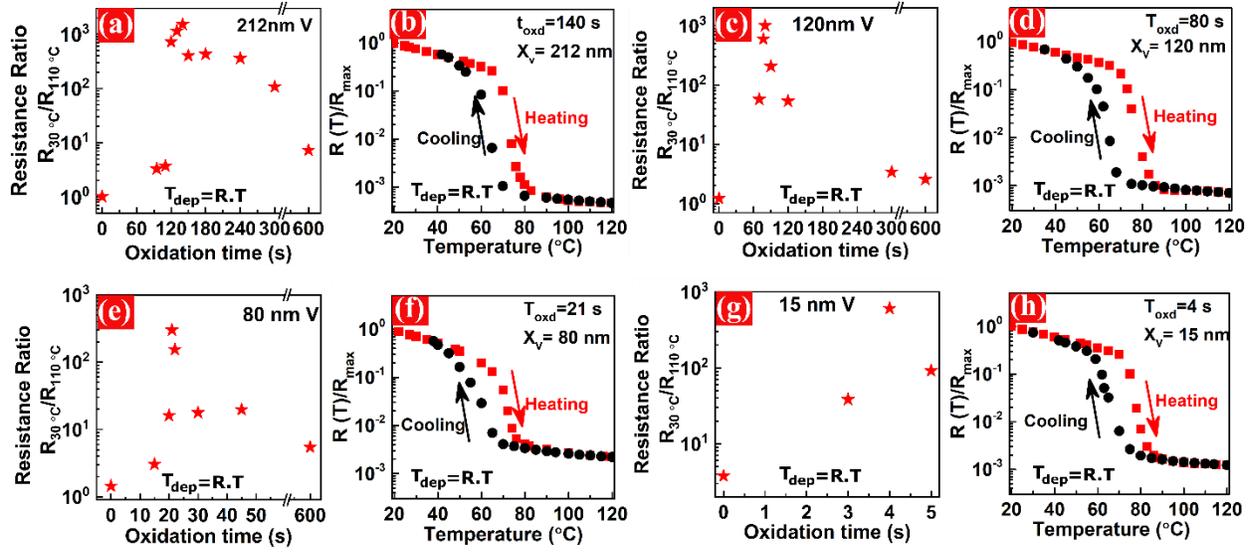

FIG. 2. The resistance switching ratio as a function of oxidation duration for Vanadium films of different thickness $X_v$ (a) 212 nm, (c) 120 nm, (e) 80 nm, and (g) 15 nm. And the reversible-resistance switching scan of temperature for $X_v$ = 212 nm and $t_{oxd}$ = 140 s (b), $X_v$ = 120 nm and $t_{oxd}$=80 s (d), $X_v$ = 80 nm and $t_{oxd}$ = 21 s (f) , $X_v$ = 30 nm and $t_{oxd}$ = 4 s (h). Resistance was beyond measurement limits for $X_V$ = 80 nm, $t_{oxd}$ = 1800 s sample, and $X_V$ = 15 nm, $t_{oxd}$ = 600 s, 1800 s samples.

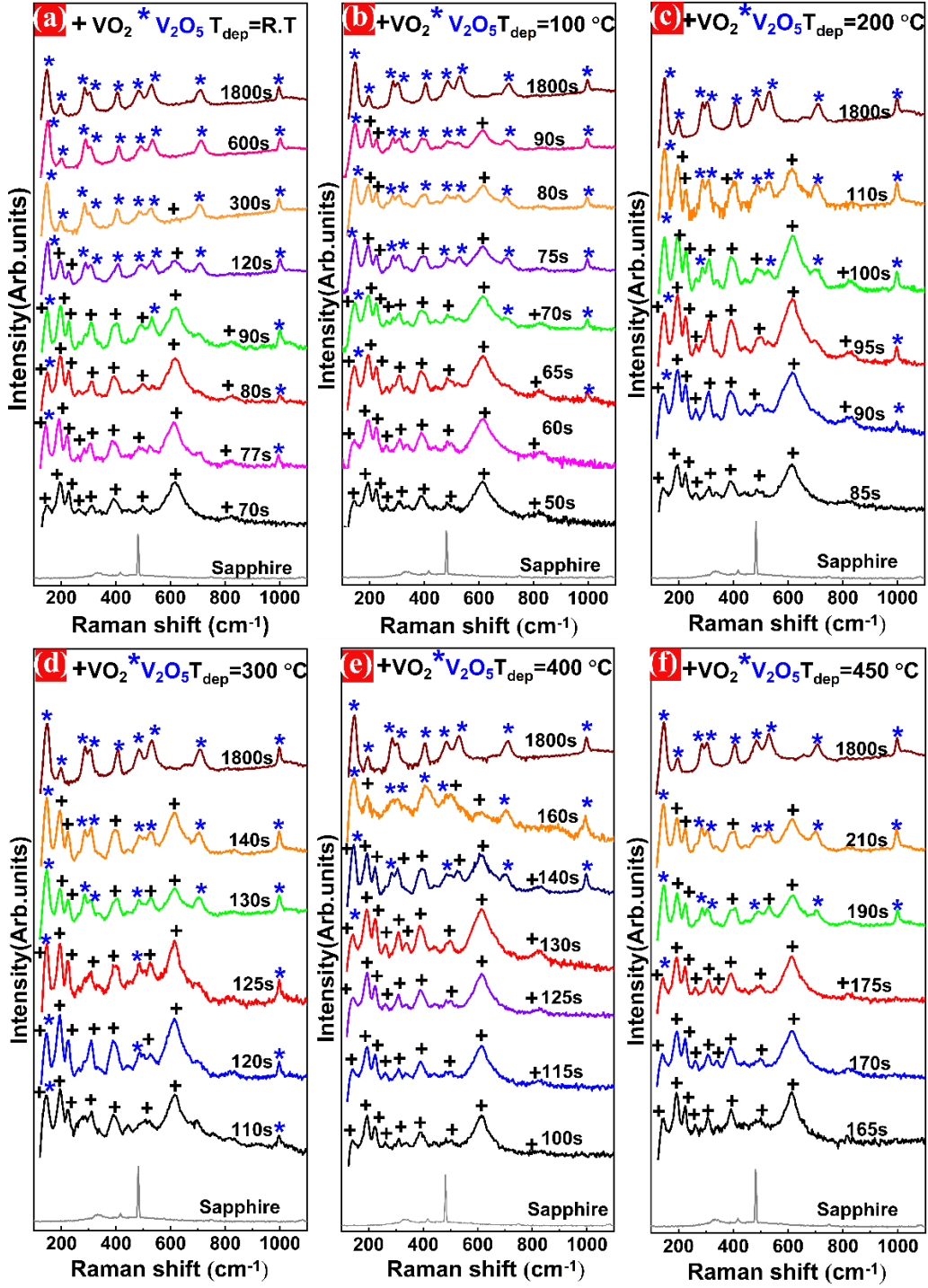

FIG. 3. Raman spectra of all oxidized Vanadium films with $X_V$ = 120 nm thickness and deposited at varying temperatures $T_{dep}$, (a) R.T, (b) 100 °C, (c) 200 °C, (d) 300 °C, (e) 400 °C, and (f) 450 °C. Reference values of the Raman peak positions for $VO_2$ and $V_2O_5$ are taken from ref. [38].

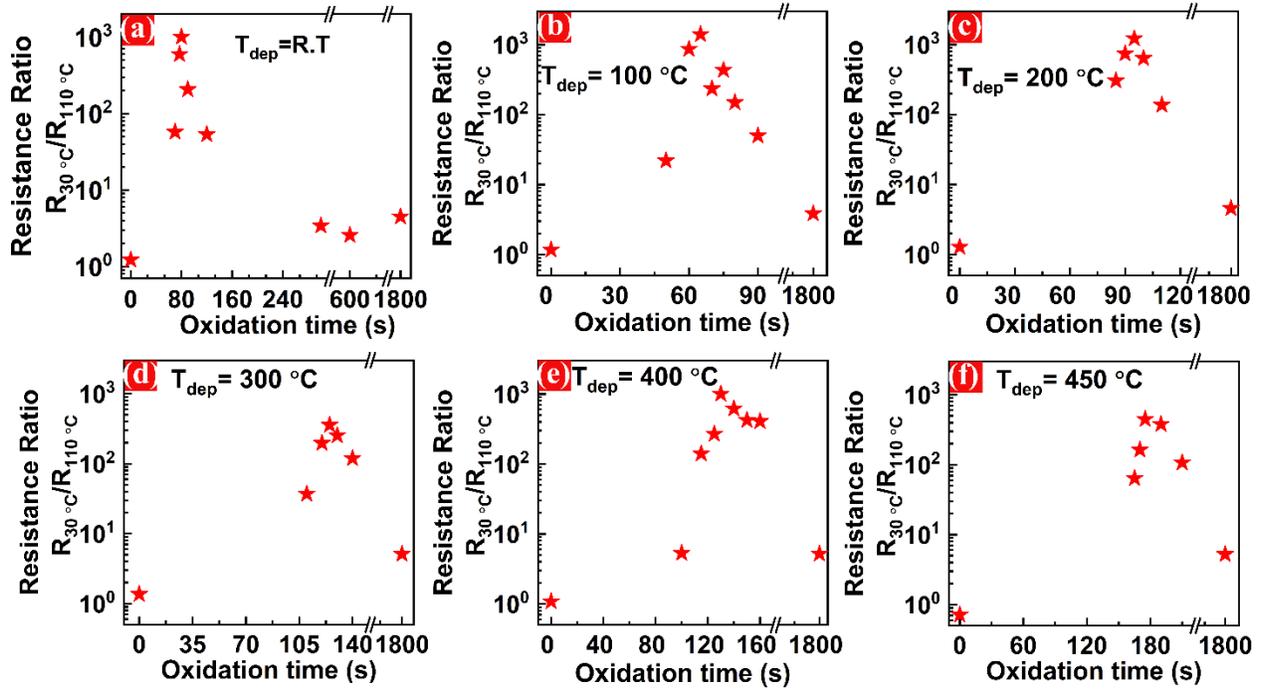

FIG. 4. The resistance switching ratio as a function of oxidation duration for $X_V = 120$ nm thick Vanadium films deposited at different temperatures $T_{dep}$, (a) R.T, (b) 100 °C, (c) 200 °C, (d) 300 °C, (e) 400 °C, and (f) 450 °C.

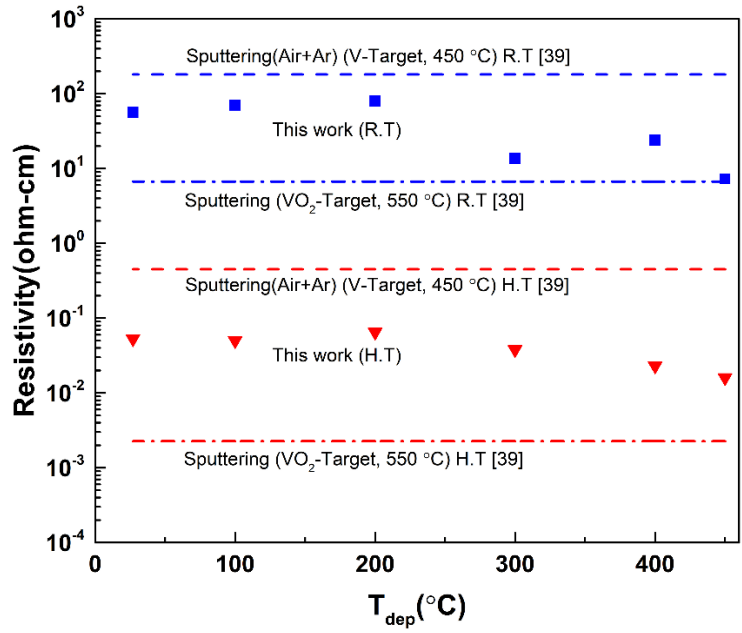

FIG. 5. Room-temperature (30 ˚C) and high-temperature (110 ˚C) resistivity of oxidized $VO_2$ films with different deposition temperature ($T_{dep}$) compared with the resistivity of $VO_2$ films prepared using reactive sputtering (V-target & Ar+air) at 450 ˚C and RF sputtering ($VO_2$- target) at 550 ˚C [39].

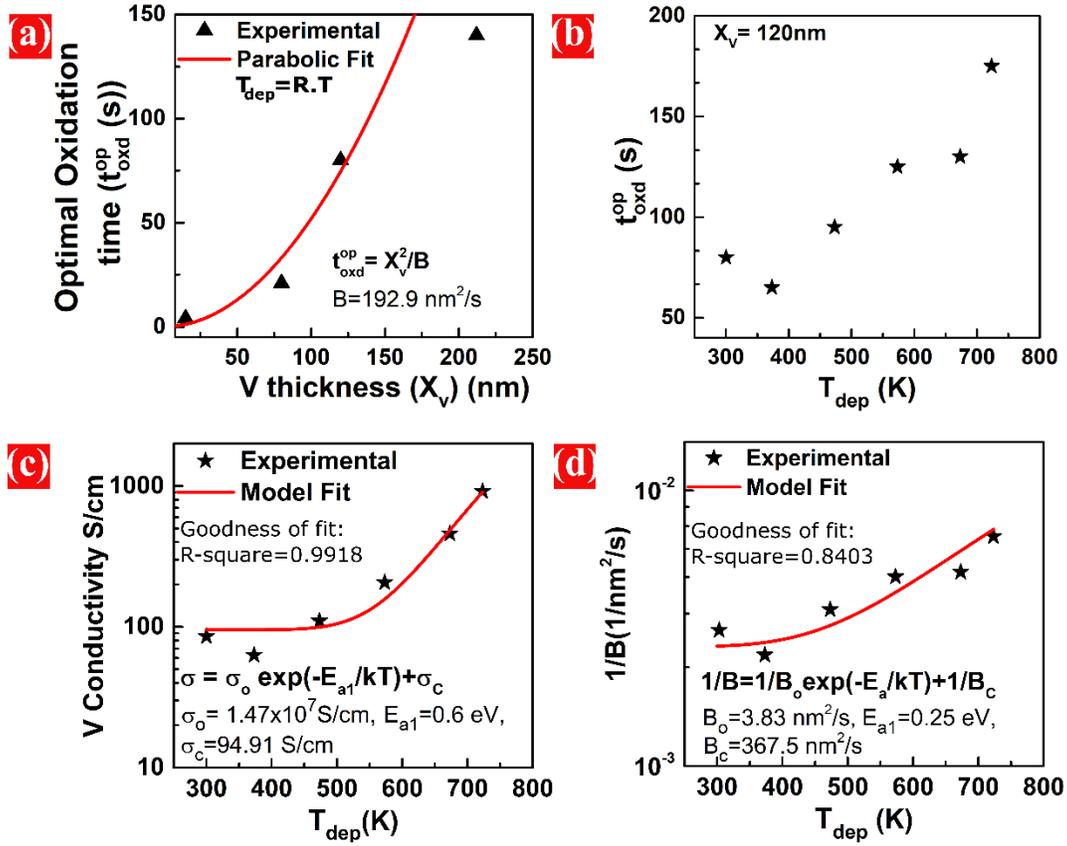

FIG. 6. (a) Obtained $t_{oxd}^{op}$ for different thickness of Vanadium $X_V$ ; Parabolic fit to the data is also shown. (b) Obtained $t_{oxd}^{op}$ as a function of V deposition temperature $T_{dep}$ for $X_V$ = 120 nm samples, (c) Measured room temperature conductivity of V films deposited at different $T_{dep}$ and model fit for the same. (d) Parabolic constant as a function of $T_{dep}$ ; model fit is also shown.